\RequirePackage{fix-cm}
\RequirePackage{fixltx2e}
\documentclass[english]{article}
\usepackage[T1]{fontenc}
\usepackage[latin9]{inputenc}
\usepackage{geometry}
\usepackage[active]{srcltx}
\usepackage{color}
\usepackage{babel}
\usepackage{array}
\usepackage{float}
\usepackage{multirow}
\usepackage{amsmath}
\usepackage{amsthm}
\usepackage{amssymb}
\usepackage{graphicx}
\usepackage{setspace}
\usepackage[numbers]{natbib}
\usepackage[unicode=true,pdfusetitle,
 bookmarks=true,bookmarksnumbered=true,bookmarksopen=true,bookmarksopenlevel=1,
 breaklinks=true,pdfborder={0 0 1},backref=false,colorlinks=true]
 {hyperref}
\hypersetup{
 linkcolor=blue, citecolor=blue, urlcolor=blue, filecolor=blue,pdfpagelayout=OneColumn, pdfnewwindow=true,pdfstartview=XYZ, plainpages=false, pdfpagelabels, hyperindex=true}
\usepackage{breakurl}

\makeatletter

\providecommand{\tabularnewline}{\\}
\floatstyle{ruled}
\newfloat{algorithm}{tbp}{loa}
\providecommand{\algorithmname}{Algorithm}
\floatname{algorithm}{\protect\algorithmname}

\numberwithin{equation}{section}
\numberwithin{figure}{section}
\numberwithin{table}{section}

\usepackage{pdflscape}
\usepackage{fancyhdr}
\usepackage{lastpage}
\usepackage{amsmath}
\usepackage{graphicx}
\usepackage{color}
\usepackage{fancybox} 
\usepackage{moreverb} 
\usepackage{listings} 
\usepackage{algorithmic}
\floatname{algorithm}{Algorithm}

\usepackage{ifpdf} 
\ifpdf 

 \IfFileExists{lmodern.sty}{\usepackage{lmodern}}{}

\fi 

\let\myTOC\tableofcontents
\renewcommand\tableofcontents{%
  \pdfbookmark[1]{\contentsname}{}
  \myTOC
}

\def\LyX{\texorpdfstring{%
  L\kern-.1667em\lower.25em\hbox{Y}\kern-.125emX\@}
  {LyX}}

\@addtoreset{footnote}{section}

\usepackage{dsfont}
\usepackage{bbm}

\theoremstyle{remark}

\makeatother

\begin{document}

\title{Local Control Regression: Improving the Least Squares\\
Monte Carlo Method for Portfolio Optimization}

\author{Rongju Zhang\thanks{Corresponding author, Email: henry.zhang@data61.csiro.au. CSIRO Data61, RiskLab Australia},
Nicolas Langren\'e\thanks{CSIRO Data61, RiskLab Australia}, Yu Tian\thanks{School of Mathematical Sciences, Monash University, Australia},
Zili Zhu\footnotemark[2], Fima Klebaner\footnotemark[3] and Kais
Hamza\footnotemark[3]}

\date{\date{}\vspace{-1cm}
}
\maketitle
\begin{abstract}
The least squares Monte Carlo algorithm has become popular for solving
portfolio optimization problems. A simple approach is to approximate
the value functions on a discrete grid of portfolio weights, then
use control regression to generalize the discrete estimates. However,
the classical global control regression can be expensive and inaccurate.
To overcome this difficulty, we introduce a local control regression
technique, combined with adaptive grids. We show that choosing a coarse
grid for local regression can produce sufficiently accurate results.\\
\vspace{-1.5mm}

\end{abstract}

\section{Introduction}

The least squares Monte Carlo (LSMC) algorithm, originally introduced
by \citep*{Carriere1996}, \citep*{Longstaff2001} and \citep*{Tsitsiklis01}
for the pricing of American options, has become a popular tool for
solving stochastic control problems in many fields, such as exotic
option pricing, inventory management, project valuation, portfolio
optimization and many others. The main strength of the LSMC algorithm
is its ability to handle multivariate stochastic state variables,
by combining the Monte Carlo simulation of the state variables with
the regression estimates of the conditional expectations in the dynamic
programming equations. 

The purpose of this paper is to improve the control discretization
approach of the LSMC algorithm for solving general portfolio optimization
problems, including general investment objective functions, general
intermediate costs and general asset dynamics. Before describing the
details of the method, we briefly review the literature on the LSMC
algorithm applied to portfolio optimization, with an emphasis on how
portfolio weights are handled and how the maximization over portfolio
weights is performed. 

The Taylor expansion method in \citep*{Brandt05} was the first attempt
to solve the portfolio optimization problems using the LSMC algorithm.
The authors determine a semi-closed form solution by deriving the
first-order condition of the Taylor series expansion of the investor\textquoteright s
future value function, see for other examples, \citep{vanBinsbergen2007},
\citep{Garlappi2010}, and \citep{Cong2017}. Another approach to
derive and solve the first-order condition is to use \citep{Glasserman2002}'s
regress-later technique, see for example, \citep{Cong2016} and \citep{Cong2016b}.
However, these two first-order condition approaches, relying on analytical
derivation, are restricted in the range of applications.

The simplest, most straightforward and most general approach to deal
with the portfolio weights in the LSMC algorithm is to discretize
them, perform one regression per discretized portfolio weight, and
compute the optimal allocation for each Monte Carlo path by naive
exhaustive search, see, for example, \citep*{Koijen09}, \citep*{Cong2017}
and \citep{Zhang2018}. Although this approach is stable and accurate,
it is computationally very expensive if a fine grid is used, especially
for multi-dimensional problems.

To improve the computational efficiency of the control discretization
approach, control regression is a possible solution. The simplest
approach is to regress state variables and portfolio weights at once.
For example, in \citep*{Denault17b} and \citep*{Denault2017}, the
value functions are regressed on the simulated exogenous state variables,
discretized wealth levels and discretized portfolio weights, and then
the first-order condition is derived and solved. In \citep{Zhang2018},
a control regression approach is implemented via the control randomization
approach of \citep*{Kharroubi2014}: the portfolio weights are randomized
and the wealth variable is correspondingly computed in the forward
simulation, afterwards the value functions are recursively regressed
on the exogenous state variables, the randomized portfolio weights
and wealth. However, as shown in \citep{Zhang2018}, this approach
may not be as accurate as the grid search approach, especially when
the time horizon is long or the payoff function is highly nonlinear.
Moreover, as discussed in \citep*{Koijen10}, regressing on state
variables and portfolio weights at once requires a very large regression,
which can make it computationally difficult. 

Another strand of control regression approach is introduced in \citep{Koijen10}.
The authors first regress the conditional expectations given in the
first-order condition for each portfolio weight on a coarse grid,
then regress the resulting regression coefficients on these portfolio
weights, obtaining a hybrid generalized conditional expectation estimate
expressed in both portfolio weight and state variables while keeping
the size of the coarse grid of portfolio weights manageable. The regression
basis of this approach is usually chosen to be polynomial (linear
in \citep{Koijen10}, quadratic in \citep*{Shen14} and \citep*{Diris15}),
in order to keep the maximization over the portfolio weights simple
(constrained optimization solver in \citep{Koijen10}, analytical
solution in \citep{Shen14} and \citep{Diris15}).

One such attempt is in \citep*{Broadie2016}. Their approach is to
sample the portfolio weights and the asset prices using low-discrepancy
Sobol sequences, then use a global regression to obtain a conditional
expectation estimate for every possible portfolio weight, and finally
solve for the optimal portfolio weight by gradient optimization.

Building upon the aforementioned approaches, we propose a hybrid framework
that adapts \citep{Zhang2018}'s combined method of control discretization
and control randomization to incorporate the control regression approach
of \citep{Broadie2016}. Our control regression approach differs from
\citep{Broadie2016} in two main aspects, namely the use of local
regression instead of global regression, and the use of adaptive refinement
grids to target the region of optimal weights instead of gradient
optimization. Our numerical experiments show that local control regression
is more accurate and more efficient than global control regression,
and that using a very coarse grid for the local control regression
can produce sufficiently accurate results.

\section{Problem Description \label{sec:portfolio-optimisation}}

We consider a finite horizon portfolio allocation problem with one
risk-free asset and $d$ risky assets available. Let $T$ be the investment
horizon. We assume the portfolio can be rebalanced at any discrete
time $t_{n}$ from the equally-spaced time grid $\Pi=\{t_{0}=0,t_{1},\ldots,t_{N}=T\}$.
Denote by $r^{f}$ the return of the risk-free asset over a single
period, by $\{\mathbf{r}_{t_{n}}\}_{0\leq n\leq N}=\{r_{t_{n}}^{i}\}_{0\leq n\leq N}^{1\leq i\leq d}$
the asset returns , by $\{\mathbf{R}_{t_{n}}\}_{0\leq n\leq N}=\{1+r_{t_{n}}^{i}\}_{0\leq n\leq N}^{1\leq i\leq d}$
the compounding factors of the asset returns, and by $\{\mathbf{S}_{t_{n}}\}_{0\leq n\leq N}=\{S_{t_{n}}^{i}\}_{0\leq n\leq N}^{1\leq i\leq d}$
the asset prices. Let $\mathbf{q}_{t_{n}}=(q_{t_{n}}^{i})_{0\leq n\leq N-1}^{1\leq i\leq d}$
describe the number of units held in each risky asset and let $\{q_{t_{n}}^{f}\}_{0\leq n\leq N-1}$
denote the amount allocated in the risk-free cash. Finally, let $\{W_{t_{n}}\}_{0\leq n\leq N}$
denote the portfolio value (wealth) process.

Let $\times$ and $\div$ denote the element-wise multiplication and
division between two vectors. The asset prices evolve as
\begin{eqnarray}
\mathbf{S}_{t_{n+1}}\ \  & = & \mathbf{S}_{t_{n}}\times\mathbf{R}_{t_{n+1}},\label{eq:price-evolution}
\end{eqnarray}
and the wealth process evolves as
\begin{equation}
W_{t_{n+1}}=W_{t_{n}}+q_{t_{n}}^{f}r^{f}+\mathbf{q}_{t_{n}}\cdot\left(\mathbf{S}_{t_{n}}\times\mathbf{r}_{t_{n+1}}\right).\label{eq:wealth-evolution}
\end{equation}

Let $\mathcal{F}=\left\{ \mathcal{F}_{t}\right\} _{0\leq t\leq T}$
be the filtration generated by all the state variables. At any rebalancing
time $t_{n}\in\Pi$, the optimization problem is
\begin{eqnarray}
v_{t_{n}}(z,w) & = & \sup_{\left\{ \boldsymbol{\alpha}_{t_{n'}}\in\mathcal{A}\right\} _{n\leq n'\leq N-1}}\mbox{\ensuremath{\mathbb{E}}}\left[U(W_{t_{N}})\left|\mathbf{Z}_{t_{n}}=z,W_{t_{n}}=w\right.\right].\label{eq:objective-function}
\end{eqnarray}
Here, $\{\mathbf{Z}_{t_{n}}\}_{0\leq n\leq N}$ is the vector of return
predictors which drive the asset price dynamics $\{\mathbf{S}_{t_{n}}\}_{0\leq n\leq N}$.
Thus, the variables $\{\mathbf{Z}_{t_{n}}\}_{0\leq n\leq N}$ form
the exogenous state variables in our problem. The investment objective
is to maximize the expected utility $\mathbb{E}[U(W_{t_{N}})]$ of
the investor's final wealth. The objective is maximized over the portfolio
allocation weights in the risky assets $\boldsymbol{\alpha}_{t_{n}}=(\alpha_{t_{n}}^{i})_{0\leq n\leq N-1}^{1\leq i\leq d}$,
with the allocation in the risk-free asset being equal to $\alpha_{t_{n}}^{f}=1-\sum_{1\leq i\leq d}\alpha_{t_{n}}^{i}$.
The relationship between portfolio weights $\boldsymbol{\alpha}_{t_{n}}$
and portfolio positions $\mathbf{q}_{t_{n}}$ is given by $\boldsymbol{\alpha}_{t_{n}}\times W_{t_{n}}=\mathbf{S}_{t_{n}}\times\mathbf{q}_{t_{n}}$.
The set $\mathcal{A}\subseteq\mathbb{R}^{d}$ of admissible strategies
is assumed time-invariant for each discrete rebalancing time. 

\section{Least Squares Monte Carlo\label{sec:lsmc}}

In this section, we describe how to use the LSMC method to solve the
dynamic portfolio optimization problem. The objective function in
Eq. \eqref{eq:objective-function} can be formulated as a discrete-time
dynamic programming principle,
\begin{eqnarray}
v_{t_{n}}(z,w) & = & \sup_{\left\{ \boldsymbol{\alpha}_{t_{n'}}\in\mathcal{A}\right\} _{n\leq n'\leq N-1}}\mbox{\ensuremath{\mathbb{E}}}\left[v_{t_{n+1}}\left(\mathbf{Z}_{t_{n+1}},W_{t_{n+1}}\right)\left|\mathbf{Z}_{t_{n}}=z,W_{t_{n}}=w\right.\right].\label{eq:dp}
\end{eqnarray}
The conventional LSMC scheme simulates the state variables forward,
then approximates the value functions recursively backwards in time
by least squares regressions. The difficulty here stems from the conditioning
wealth variable $W_{t_{n}}$, because it requires the information
of all the previous portfolio decisions $\{\boldsymbol{\alpha}_{t_{n'}}\}_{0\leq n'\leq n}$
which are not known yet in the backward dynamic programming loop.
To overcome this difficulty, a popular approach is to approximate
the conditional expectations in Eq. \eqref{eq:objective-function}
on a discrete grid of wealth levels, and then interpolate them on
these discrete wealth levels, see for example, \citep{Brandt05},
\citep*{Denault17b} and \citep*{Denault2017}. 

The discretization approach can be computationally demanding if a
fine grid is required, especially when there are multiple endogenous
state variables. For example, when market impact is accounted for,
the asset prices become endogenous state variables, which can make
the discretization approach computationally infeasible. 

Instead, we use a control randomization approach that simulates random
portfolio weights $\{\tilde{\boldsymbol{\alpha}}_{t_{n}}^{m}\}_{0\leq n\leq N-1}^{1\leq m\leq M}$
in the forward loop in addition to the return predictors $\{\mathbf{Z}_{t_{n}}^{m}\}_{0\leq n\leq N}^{1\leq m\leq M}$,
the asset returns $\{\mathbf{r}_{t_{n}}^{m}\}_{0\leq n\leq N}^{1\leq m\leq M}$,
and the asset prices $\{\mathbf{S}_{t_{n}}^{m}\}_{0\leq n\leq N}^{1\leq m\leq M}$.
Simulations for the wealth variable $\{\tilde{W}{}_{t_{n}}^{m}\}_{0\leq n\leq N}^{1\leq m\leq M}$
can be subsequently computed using the randomized portfolio weights.
These portfolio weights are uniformly drawn from the admissible set
$\mathcal{A}$, so as to ensure that the simulated sample covers the
whole space of possible portfolio weights.

After all the state variables have been simulated, the next task is
to approximate the conditional expectations in the dynamic programming
formula in Eq. \eqref{eq:dp} by regression. A possible approach,
called control regression, is to regress on the randomized portfolio
weights as well as the return predictors, and then compute the optimal
decisions by solving the first-order conditions. Such a control regression
scheme, combined with control randomization, is analyzed in \citep{Kharroubi2014},
and an application of this approach for solving dynamic portfolio
optimization problem is given in \citep{Zhang2018}.

However, regressing on the portfolio weights and the state variables
at once requires a very large regression, which can make it computationally
difficult, especially for high dimensional portfolios.

Instead, we resort to control discretization, which has been shown
to be more accurate than control regression in \citep{Zhang2018}.
The control space is discretized as $\mathcal{A}\approx\mathcal{A}^{\text{disc}}=\{\mathbf{a}_{1},...,\mathbf{a}_{J}\}$.
At time $t_{n}$, assuming the mapping $\hat{v}_{t_{n+1}}:\left(z,w\right)\mapsto\hat{v}_{t_{n+1}}\left(z,w\right)$
has been estimated, the dynamic programming principle in Eq. \eqref{eq:dp}
becomes
\[
v_{t_{n}}(z,w)=\max_{\mathbf{a}_{j}\in\mathcal{A}^{\text{disc}}}\mathbb{E}\left[\hat{v}_{t_{n+1}}\left(\mathbf{Z}_{t_{n+1}},W_{t_{n+1}}\right)\left|\mathbf{Z}_{t_{n}}=z,W_{t_{n}}=w,\boldsymbol{\alpha}_{t_{n}}=\mathbf{a}_{j}\right.\right]=:\max_{\mathbf{a}_{j}\in\mathcal{A}^{\text{disc}}}\text{CV}_{t_{n}}^{j}\left(z,w\right),
\]
where $\text{CV}_{t_{n}}^{j}$ denotes the continuation value function
of choosing the portfolio weight $\mathbf{a}_{j}$ at time $t_{n}$.
To evaluate $\text{CV}_{t_{n}}^{j}$ for each $\mathbf{a}_{j}\in\mathcal{A}^{\text{disc}}$,
we first update the portfolio weights $\boldsymbol{\alpha}_{t_{n}}^{m}=\mathbf{a}_{j}$
for each Monte Carlo path $m=1,...,M$. Then, the wealth variables
at time $t_{n+1}$ can be recomputed based on the portfolio weight
$\mathbf{a}_{j}$,
\begin{eqnarray*}
\mathbf{q}_{t_{n}}^{m,j} & = & \mathbf{a}_{j}\times\tilde{W}_{t_{n}}^{m}\div\mathbf{S}_{t_{n}}^{m}\\
\hat{W}_{t_{n+1}}^{m,(n,j)} & = & \tilde{W}_{t_{n}}^{m}+q_{t_{n}}^{f,m,j}r^{f}+\mathbf{q}_{t_{n}}^{m,j}\cdot\left(\mathbf{S}_{t_{n}}^{m}\times\mathbf{r}_{t_{n+1}}^{m}\right),
\end{eqnarray*}
where $\hat{W}_{t_{n+1}}^{m,(n,j)}$ is the recomputed wealth at time
$t_{n+1}$ conditioning on the information at time $t_{n}$ at the
$j^{\text{th}}$ node of the control grid. Let $\{\psi_{k}\left(z,w\right)\}_{1\leq k\leq K}$
be the basis functions of the state variables for the regression.
For each portfolio decision $\mathbf{a}_{j}\in\mathcal{A}^{\text{disc}}$,
the corresponding $\text{CV}_{t_{n}}^{j}$ is approximated by least
squares minimization:
\begin{eqnarray*}
\left\{ \hat{\beta}_{k,t_{n}}^{j}\right\} _{1\leq k\leq K} & = & \arg\min_{\left\{ \beta_{k}\right\} _{1\leq k\leq K}}\sum_{m=1}^{M}\left(\hat{v}_{t_{n+1}}\left(\mathbf{Z}_{t_{n+1}}^{m},\hat{W}_{t_{n+1}}^{m,(n,j)}\right)-\sum_{k=1}^{K}\beta_{k}\psi_{k}\left(\mathbf{Z}_{t_{n}}^{m},\tilde{W}{}_{t_{n}}^{m}\right)\right)^{2}.
\end{eqnarray*}
Finally, $\text{CV}_{t_{n}}^{j}$ is parametrized as 
\begin{eqnarray*}
\hat{\text{CV}}_{t_{n}}^{j}\left(z,w\right) & = & \sum_{k=1}^{K}\beta_{k,t_{n}}^{j}\psi_{k}\left(z,w\right),
\end{eqnarray*}
and the optimal decision policy is
\begin{eqnarray*}
\hat{\boldsymbol{\alpha}}_{t_{n}}\left(z,w\right)=\arg\max_{\mathbf{a}_{j}\in\mathcal{A}^{\text{disc}}}\hat{\text{CV}}_{t_{n}}^{j}\left(z,w\right), &  & \hat{v}\left(z,w\right)=\max_{\mathbf{a}_{j}\in\mathcal{A}^{\text{disc}}}\hat{\text{CV}}_{t_{n}}^{j}\left(z,w\right),
\end{eqnarray*}
where the maximization is performed by an exhaustive grid search. 

As tested in \citep{Zhang2018}, the control discretization approach
is numerically very stable, even when transaction costs and price
impact are taken into account. However, the exhaustive grid search
is computationally very expensive, especially for high dimensional
portfolios. In the next section, we will introduce a local control
regression method to reduce the computational burden of control discretization.

\section{Control Regression and Maximization\label{sec:local-regression-adaptive grids}}

After the value functions have been estimated on the discrete grid
$\mathcal{A}^{\text{disc}}$, the next step is to use control regression
to project these discrete estimates onto a continuous space of control.
This section describes in detail how to use control regression to
improve the estimates. For notional convenience, we denote by $\hat{\text{CV}}^{j}=\hat{\text{CV}}^{j}\left(z,w\right)$
the estimated continuation value of choosing the portfolio weight
$\mathbf{a}_{j}$ at an arbitrary time and for an arbitrary input
of state variables $(z,w)$. Suppose we have already implemented the
grid search approach to estimate the conditional values $\hat{\text{CV}}^{j}$
on a coarse discrete grid $\mathcal{A}^{\text{disc}}$, and have estimated
the optimal control by the following equation
\begin{equation}
\hat{\boldsymbol{\alpha}}=\arg\max_{\mathbf{a}_{j}\in\mathcal{A}^{\text{disc}}}\hat{\text{CV}}^{j}.\label{eq:discrete-control}
\end{equation}

\begin{figure}
\begin{centering}
\caption{Global control regression\label{fig:global_regression}}
\par\end{centering}
\medskip
\centering{}%
\begin{minipage}[t]{1\columnwidth}%
\begin{center}
\includegraphics[scale=0.12]{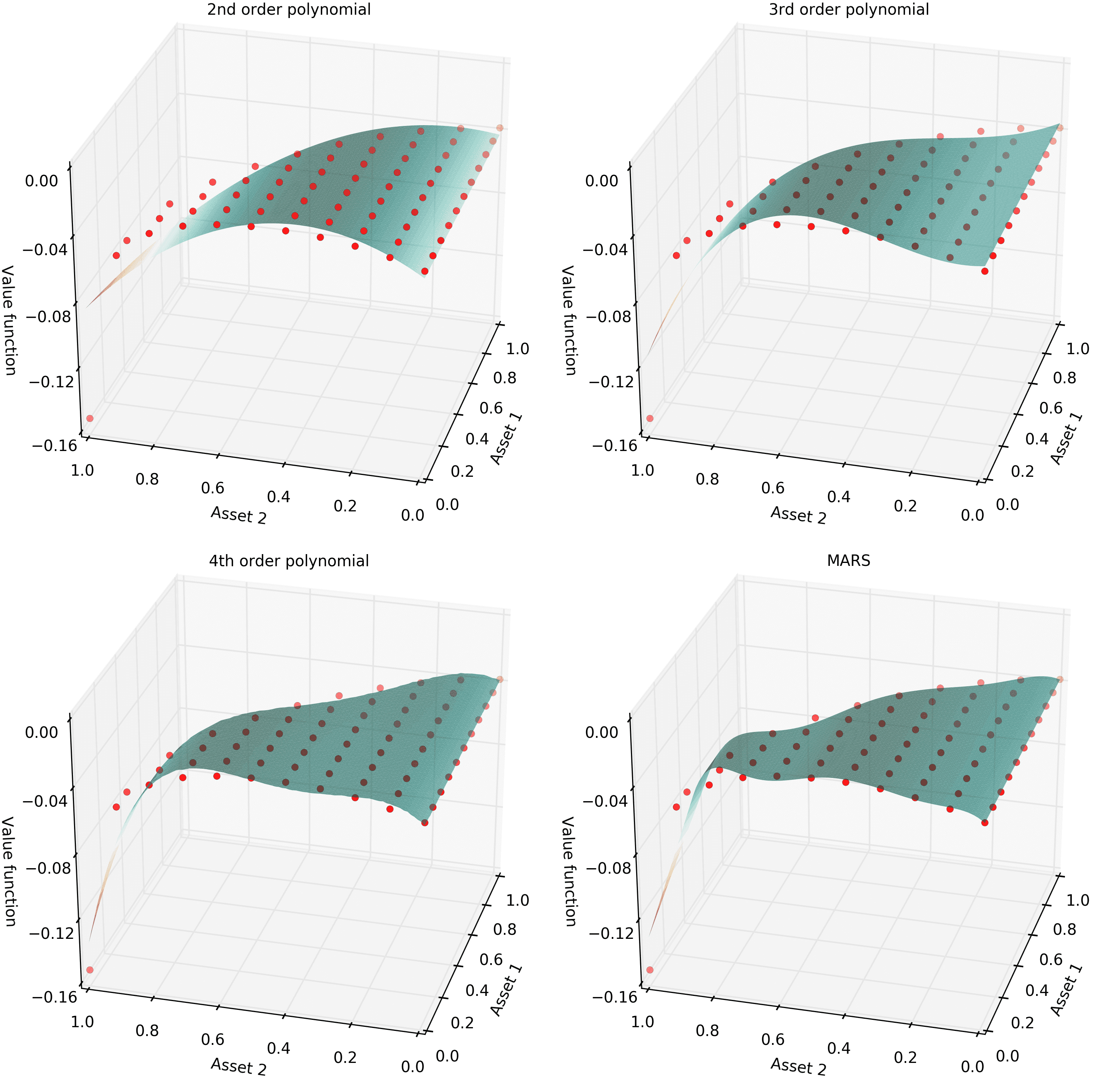}
\par\end{center}%
\end{minipage}
\end{figure}

The usual approach to generalize these conditional value estimates
on the discrete grid $\mathcal{A}^{\text{disc}}$ is to use global
regression, see for example \citep{Koijen10} and \citep{Broadie2016}.
However, it can be hard to find a regression basis that fits the global
grid well, even when using advanced regression techniques. As an example,
Figure \ref{fig:global_regression} illustrates the global regression
approach for a portfolio with risk-free cash and two risky assets
($d=2$). Moreover, even if adequate regression basis can be found,
the global regression can be computationally very expensive. In fact,
it is unnecessary to project the discrete value functions onto the
whole global control space, because what matters for maximizing the
value functions is the accuracy on the small local region containing
the optimum.
\begin{figure}
\begin{centering}
\caption{Local control regression\label{fig:local_regression}}
\par\end{centering}
\medskip
\begin{centering}
\hspace{4.5em}%
\begin{minipage}[t]{0.48\columnwidth}%
\includegraphics[scale=0.24]{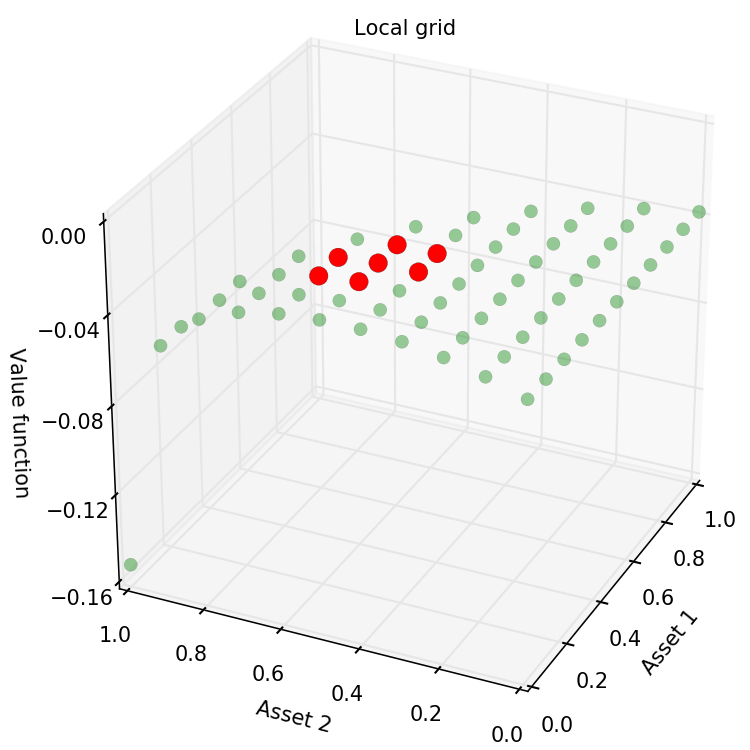}%
\end{minipage}\hspace{-4.5em}%
\begin{minipage}[t]{0.48\columnwidth}%
\includegraphics[scale=0.24]{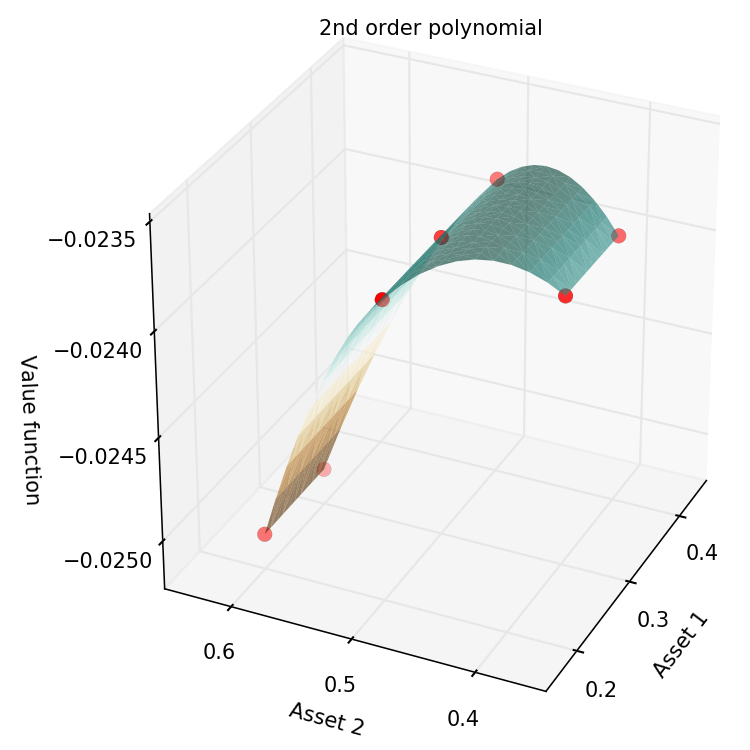}%
\end{minipage}
\par\end{centering}
{\footnotesize{}The left panel highlights the local control grid (in
red) out of the global grid (in green) and the right panel shows the
fitted surface on the selected local grid. }{\footnotesize\par}
\end{figure}

\subsection{Local control regression\label{subsec:local-regression}}

In order to improve the numerical accuracy and efficiency of control
regression, we propose to perform a local regression around the optimal
estimate $\hat{\boldsymbol{\alpha}}$. We construct a local grid $\mathcal{L}^{\text{disc}}(\hat{\boldsymbol{\alpha}})\subset\mathcal{A}^{\text{disc}}$
that contains the neighbors of $\hat{\boldsymbol{\alpha}}$ :
\begin{equation}
\mathcal{L}^{\text{disc}}\left(\hat{\boldsymbol{\alpha}}\right):=\left\{ \mathbf{a}\in\mathcal{A}^{\text{disc}}:\left\Vert \hat{\boldsymbol{\alpha}}-\mathbf{a}\right\Vert _{\infty}\leq\delta\right\} ,
\end{equation}
where $\delta$ is the mesh size of the discrete grid $\mathcal{A}^{\text{disc}}$.
Then, we regress the estimated conditional values $\hat{\text{CV}}^{j}$
with respect to the points in the local grid $\mathcal{L}^{\text{disc}}(\hat{\boldsymbol{\alpha}})$,
obtaining the parametric conditional value estimates $\hat{\Phi}\left(\mathbf{a}\right)$
over the whole continuous local control hypercube
\begin{equation}
\mathcal{L}^{\text{cont}}\left(\hat{\boldsymbol{\alpha}}\right):=\left\{ \mathbf{a}\in\mathcal{A}:\left\Vert \hat{\boldsymbol{\alpha}}-\mathbf{a}\right\Vert _{\infty}\leq\delta\right\} .
\end{equation}
The main advantage of local regression is a better goodness-of-fit,
as the local grid $\mathcal{L}^{\text{disc}}\left(\hat{\boldsymbol{\alpha}}\right)$
contains only very few points and are therefore easier to fit even
with simple bases such as local second-order polynomial bases. In
addition, a simple second-order polynomial is a sensible candidate
for such a local parametric regression, as the local grid only contains
a maximum of three points in each dimension and the maximum lies at
either the middle point or the boundary of the local coarse grid.
Furthermore, to guarantee the robustness of local regression, we perform
local Ridge regressions with second-order polynomial bases. Figure
\ref{fig:local_regression} illustrates the local control regression
approach on the same example presented in Figure \ref{fig:global_regression}. 

Now that continuation value estimates have been obtained over the
whole local hypercube $\mathcal{L}^{\text{cont}}(\hat{\boldsymbol{\alpha}})$,
we turn to the problem of estimating the optimal control over $\mathcal{L}^{\text{cont}}(\hat{\boldsymbol{\alpha}})$,
i.e.,
\begin{eqnarray}
\hat{\boldsymbol{\alpha}}^{*} & = & \arg\max_{\mathbf{a}\in\mathcal{L}^{\text{cont}}\left(\hat{\boldsymbol{\alpha}}\right)}\hat{\Phi}\left(\boldsymbol{a}\right).\label{eq:improved-control}
\end{eqnarray}

\subsection{Maximization by adaptive grids}

The next step is to improve the discrete optimal control estimate
in Eq. \eqref{eq:discrete-control} by solving the continuous equation
in Eq. \eqref{eq:improved-control}. Classical approaches to do so
are to use gradient optimization which requires either analytically
or numerically compute the gradients. Instead, we perform the maximization
using a gradient-free adaptive refinement strategy.

Here is a simple illustration of this adaptive grid method. Suppose
we are given a portfolio of a risk-free asset and a risky asset. We
denote by $\{\mathcal{A}_{p}\}_{0\leq p\leq P}$ the sequence of adaptive
grids and denote by $\{\delta_{p}\}_{0\leq p\leq P}$ the sequence
as adaptive mesh sizes such that $\delta_{p+1}=\delta_{p}/2$. We
define the initial adaptive grid as $\mathcal{A}_{0}=\mathcal{A}^{\text{disc}}$
and define the initial control mesh size as $\delta_{0}=\delta$.
Suppose the optimal control $\hat{\alpha}_{0}$ has been estimated
on the space $\mathcal{A}_{0}=\mathcal{A}^{\text{disc}}$. Then, the
updated adaptive grid is defined as $\mathcal{A}_{1}=\{\hat{\alpha}_{0}-\delta_{0}/2,\hat{\alpha}_{0},\hat{\alpha}_{0}+\delta_{0}/2\}$
and the updated estimate of optimal control is given by $\hat{\alpha}_{1}=\arg\max_{a\in\mathcal{A}_{1}}\hat{\Phi}\left(a\right)$.
Iteratively, the adaptive grid will be $\mathcal{A}_{p}=\{\hat{\alpha}_{p-1}-\delta/2^{p},\hat{\alpha}_{p-/1},\hat{\alpha}_{p-1}+\delta/2^{p}\}$.
For every refined adaptive grid, there is a maximum of two new points
in each dimension of the control. This hierarchical structure guarantees
a limited number of operations will be performed in the algorithm.
Below is an example of our adaptive grid refinement strategy starting
with a $\delta=1/4$ control mesh for $\mathcal{A}^{\text{disc}}$:
\begin{eqnarray*}
 & \mathcal{A}^{\text{d}}= & \mathcal{A}_{0}=\{0,\,0.2500,\,0.5000,\,0.7500,\,1\}\\
\text{\text{Suppose }}\hat{\alpha}_{0}=\arg\max_{a\in\mathcal{A}_{0}}\hat{\Phi}\left(a\right)=0.5000, & \text{then} & \mathcal{A}_{1}=\{0.3750,\,0.5000,\,0.6250\}\\
\text{Suppose }\hat{\alpha}_{1}=\arg\max_{a\in\mathcal{A}_{1}}\hat{\Phi}\left(a\right)=0.3750, & \text{then} & \mathcal{A}_{2}=\{0.3125,\,0.3750,\,0.4375\}\\
\text{Suppose }\hat{\alpha}_{2}=\arg\max_{a\in\mathcal{A}_{2}}\hat{\Phi}\left(a\right)=0.3125, & \text{then} & \mathcal{A}_{3}=\{0.28125,\,0.3125,\,0.34375\}\\
\text{Suppose }\hat{\alpha}_{3}=\arg\max_{a\in\mathcal{A}_{3}}\hat{\Phi}\left(a\right)=0.3125, & \text{then} & \mathcal{A}_{4}=\{0.296875,\,0.3125,\,0.32815\}\\
 & \vdots\\
\text{Finally }\hat{\alpha}^{*}:=\arg\max_{a\in\mathcal{A}_{P}}\hat{\Phi}\left(a\right)
\end{eqnarray*}
One can see that it takes only five iterations to reach a control
mesh precision higher than $1/100$, and only a maximum of two extra
evaluation points are required at each iteration. This adaptive grid
method can be described as a modified multivariate bisection search
for the local maximum of a function. The combination of local regression
and adaptive grids is described in Algorithm \ref{algo:adaptive}.
In our experience, both gradient optimization and adaptive grids can
reach the optimum but the latter is more efficient. 

\begin{algorithm}[H]
\begin{algorithmic}[1]

\STATE \textbf{Input:} $\left\{ \hat{\text{CV}}^{j}\right\} _{1\leq j\leq J}$ 

\STATE \textbf{Result:} $\hat{\boldsymbol{\alpha}}^{*}$ 

\STATE Find the optimal decision on the coarse grid by grid searching:

\vspace{-0.5em}
\[
\hat{\boldsymbol{\alpha}}_{0}=\arg\max_{\mathbf{a}_{j}\in\mathcal{A}^{\text{disc}}}\hat{\text{CV}}{}^{j}
\]

\vspace{-0.5em}

\STATE Construct a local grid in the neighbourhood of $\hat{\boldsymbol{\alpha}}$:

\vspace{-0.5em}
\[
\mathcal{L}^{\text{disc}}\left(\hat{\boldsymbol{\alpha}}_{0}\right):=\left\{ \mathbf{a}\in\mathcal{A}^{\text{disc}}:\left\Vert \hat{\boldsymbol{\alpha}}_{0}-\mathbf{a}\right\Vert _{\infty}\leq\delta\right\} 
\]

\vspace{-0.5em}

\STATE Regress $\left\{ \hat{\text{CV}}^{j}\right\} _{\mathbf{a}_{j}\in\mathcal{L}^{\text{disc}}\left(\hat{\boldsymbol{\alpha}}_{0}\right)}$
on $\mathcal{L}^{\text{disc}}\left(\hat{\boldsymbol{\alpha}}_{0}\right)$,
obtaining a parametric shape $\hat{\Phi}\left(\boldsymbol{\alpha}_{0}\right)$
on 

\vspace{-0.5em}
\[
\mathcal{L}^{\text{cont}}\left(\hat{\boldsymbol{\alpha}}_{0}\right):=\left\{ \mathbf{a}\in\mathcal{A}:\left\Vert \hat{\boldsymbol{\alpha}}_{0}-\mathbf{a}\right\Vert _{\infty}\leq\delta\right\} .
\]

\STATE Set the initial adaptive grid $\mathcal{A}_{0}=\mathcal{L}^{\text{disc}}\left(\hat{\boldsymbol{\alpha}}_{0}\right)$

\FORALL{$p=1,\ldots,P$}

\STATE Update adaptive grid: $\mathcal{A}_{p}:=\left\{ \mathbf{a}\in\mathcal{L}^{\text{cont}}\left(\hat{\boldsymbol{\alpha}}_{0}\right):\mathbf{a}=\hat{\boldsymbol{\alpha}}_{p-1}\pm\delta/2^{p}\right\} \cup\left\{ \hat{\boldsymbol{\alpha}}_{p-1}\right\} $

\STATE Update adaptive control estimate: $\hat{\boldsymbol{\alpha}}_{p}=\arg{\displaystyle \max_{\mathbf{a}\in\mathcal{A}_{p}}}\hat{\Phi}\left(\mathbf{a}\right)$

\ENDFOR

\STATE $\hat{\boldsymbol{\alpha}}^{*}:=\hat{\boldsymbol{\alpha}}_{P}$

\end{algorithmic}

\caption{Local regression and adaptive grids\label{algo:adaptive}}
\end{algorithm}

\section{Numerical Experiments\label{sec:Application}}

In this section, we validate the numerical method proposed in the
previous sections by applying it to a multiperiod portfolio optimization
problem as formulated in Section \ref{sec:portfolio-optimisation}.

We obtain monthly close prices from October 2007 to January 2016 of
13 financial assets (ETF tickers: BND, SPY, EFA, EEM, GLD, BWX, SLV,
USO, UUP, FXE, FXY, and FXA) from Yahoo finance. We calibrate a first-order
vector autoregressive model to log-returns ($\log S_{t}-\log S_{t-1}$)
to obtain the return dynamics. We fix the annual interest rate on
the cash account to $4.5\%$. We consider a three-dimensional portfolio
optimization with the cash component, BND and SPY. The rest of the
11 financial assets are used as exogenous return predictors for BND
and SPY. In the numerical tests, we use a sample of $M=10^{4}$ Monte
Carlo paths.

We assume transaction costs of $0.3\%$ proportional to the portfolio
turnover. For liquidity costs and market impact which depend on the
transaction volume, we describe the price movement during the transaction
by the marginal supply-demand curve (MSDC) calibrated by \citep*{Tian2013}
to the European medium- and large-cap equities. The MSDC reads
\begin{equation}
\textrm{MSDC}\left(\Delta q\right)=\begin{cases}
S_{\text{A}}e^{k\sqrt{\left|\Delta q\right|}} & \mathrm{when}\,q>0\,\text{(position increase)}\\
S_{\text{B}}e^{-k\sqrt{\left|\Delta q\right|}} & \mathrm{when}\,q<0\,\mbox{\text{(position reduction})}
\end{cases},\label{eq:msdcliquid-1}
\end{equation}
where $S_{\mathrm{B}}$ denotes the best bid price, $S_{\mathrm{A}}$
denotes the best ask price, $\Delta q$ denotes the position change
of the investor at time $t$ and $k\in\mathbb{R}^{+}$ is called the
liquidity risk factor. By integrating with respect to $\Delta q$
, we obtain an explicit parametric form for the liquidity costs,
\[
\textrm{LC}\left(\Delta q\right)=S_{\text{A}}\left(\lambda\left(\Delta q\right)-\Delta q\right)\mathbbm{1}\left\{ \Delta q>0\right\} +S_{\text{B}}\left(q-\lambda\left(\Delta q\right)\right)\mathbbm{1}\left\{ \Delta q<0\right\} \,,
\]
where $\lambda(\Delta q)=\frac{2}{k^{2}}(\mathrm{sign}(\Delta q)k\sqrt{|\Delta q|}e^{-\mathrm{sign}(q)k\sqrt{|\Delta q|}}+e^{-\mathrm{sign}(\Delta q)k\sqrt{|\Delta q|}}-1).$
Following the calibration results from \citep{Tian2013}, we set $k=8\times10^{-6}$.
The calibrated bid-ask spread $S_{\text{A}}-S_{\text{B}}$ is about
$10^{-5}$, which, according to our tests, is small enough to have
virtually no impact on the optimal portfolio allocation. For convenience,
we simply set $S_{\text{B}}=S_{\text{A}}$. The permanent market impact
can be deemed proportional to the temporary peak of the MSDC. We assume
$2/3$ for the proportional rate for the market impact. For the details
of how to incorporate these switching costs into the LSMC method,
we refer to \citep{Zhang2018}.

We use the constant relative risk aversion (CRRA) utility, i.e., $U(w)=w^{1-\gamma}/(1-\gamma)$.
for the optimization objective. We report our numerical results in
terms of monthly adjusted certainty equivalent returns (CER) given
by $\mathrm{CER}=U^{-1}(\mathbb{E}[U(W_{T})])^{\frac{1}{T}}-1\approx U^{-1}(\frac{1}{M}\sum_{m=1}^{M}U(W_{T}^{m}))^{\frac{1}{T}}-1.$
The magnitude of monthly returns is usually less than one percent,
thus we report CER in basis points to make comparisons easier. 

\subsection{Local regression v.s. global regression\label{subsec:Local-regression}}

As explained in Section \ref{subsec:local-regression}, the optimal
portfolio allocation computed on a discrete grid can be improved by
control regression. Here, we compare local control regression and
global control regression in terms of accuracy and efficiency, given
that both use adaptive grids to find the optimum. Table \ref{tab:regression}
and Table \ref{tab:regression_runtime} compare local regression to
different types of global bases. Our results show that local regression
does improve both accuracy and efficiency. The outlier value $-12.3$
in Table \ref{tab:regression} for $4^{\text{th}}$-order global polynomial
regression and $\delta=1/8$ suggests that high-order polynomial bases
may produce unreliable results. 

\begin{table}[H]
\begin{singlespace}
\begin{centering}
\caption{CER for different regression methods and different mesh sizes{\footnotesize{}\label{tab:regression}}}
\smallskip
\par\end{centering}
\begin{centering}
{\footnotesize{}}%
\begin{tabular}{ccccccccccc}
 &  & {\footnotesize{}$2^{\text{nd}}$ Local} &  & {\footnotesize{}$2^{\text{nd}}$ Global} &  & {\footnotesize{}$3^{\text{rd}}$ Global} &  & {\footnotesize{}$4^{\text{th}}$ Global} &  & {\footnotesize{}MARS Global}\tabularnewline
\hline 
{\footnotesize{}$\delta=1/2$} &  & {\footnotesize{}$64.9$} &  & {\footnotesize{}$65.0$} &  & {\footnotesize{}$64.9$} &  & {\footnotesize{}$64.8$} &  & {\footnotesize{}$63.2$}\tabularnewline
{\footnotesize{}$\delta=1/4$} &  & {\footnotesize{}$65.0$} &  & {\footnotesize{}$64.8$} &  & {\footnotesize{}$65.0$} &  & {\footnotesize{}$64.9$} &  & {\footnotesize{}$64.5$}\tabularnewline
{\footnotesize{}$\delta=1/8$} &  & {\footnotesize{}$65.1$} &  & {\footnotesize{}$64.9$} &  & {\footnotesize{}$64.5$} &  & {\footnotesize{}$-12.3$} &  & {\footnotesize{}$65.0$}\tabularnewline
{\footnotesize{}$\delta=1/16$} &  & {\footnotesize{}$65.2$} &  & {\footnotesize{}$64.8$} &  & {\footnotesize{}$64.9$} &  & {\footnotesize{}$65.0$} &  & {\footnotesize{}$65.0$}\tabularnewline
{\footnotesize{}$\delta=1/32$} &  & {\footnotesize{}$65.2$} &  & {\footnotesize{}$64.8$} &  & {\footnotesize{}$64.7$} &  & {\footnotesize{}$65.0$} &  & {\footnotesize{}$64.8$}\tabularnewline
\hline 
\end{tabular}{\footnotesize\par}
\par\end{centering}
\end{singlespace}
\smallskip

{\footnotesize{}This table compares the lower bound of the truncated
VFI scheme for the monthly adjusted CER (in basis points) using different
sizes of control mesh and different regression bases. A CRRA investment
style with $\gamma=10$ over a 6-month horizon ($N=6$ months) is
considered.}{\footnotesize\par}
\end{table}

\begin{table}[H]
\begin{singlespace}
\begin{centering}
\caption{Computational runtime for different regression methods and different
mesh sizes{\footnotesize{}\label{tab:regression_runtime}}}
\smallskip
\par\end{centering}
\begin{centering}
\begin{tabular}{cclllllllll}
\multirow{1}{*}{} &  & {\footnotesize{}$2^{\text{nd}}$ Local} &  & {\footnotesize{}$2^{\text{nd}}$ Global} &  & {\footnotesize{}$3^{\text{rd}}$ Global} &  & {\footnotesize{}$4^{\text{th}}$ Global} &  & {\footnotesize{}MARS Global}\tabularnewline
\hline 
{\footnotesize{}$\delta=1/2$} &  & {\footnotesize{}$\tau^{a}=43$ secs} &  & {\footnotesize{}$1.2\times\tau^{a}$} &  & {\footnotesize{}$1.6\times\tau^{a}$} &  & {\footnotesize{}$2.6\times\tau^{a}$} &  & {\footnotesize{}$3.6\times\tau^{a}$}\tabularnewline
{\footnotesize{}$\delta=1/4$} &  & {\footnotesize{}$\tau^{b}=1.5$ mins} &  & {\footnotesize{}$1.4\times\tau^{b}$} &  & {\footnotesize{}$1.9\times\tau^{b}$} &  & {\footnotesize{}$3.0\times\tau^{b}$} &  & {\footnotesize{}$10.5\times\tau^{b}$}\tabularnewline
{\footnotesize{}$\delta=1/8$} &  & {\footnotesize{}$\tau^{c}=5.2$ mins} &  & {\footnotesize{}$1.8\times\tau^{c}$} &  & {\footnotesize{}$2.3\times\tau^{c}$} &  & {\footnotesize{}$7.0\times\tau^{c}$} &  & {\footnotesize{}$15.6\times\tau^{c}$}\tabularnewline
{\footnotesize{}$\delta=1/16$} &  & {\footnotesize{}$\tau^{d}=24.1$ mins} &  & {\footnotesize{}$2.8\times\tau^{d}$} &  & {\footnotesize{}$3.9\times\tau^{d}$} &  & {\footnotesize{}$9.3\times\tau^{d}$} &  & {\footnotesize{}$22.3\times\tau^{d}$}\tabularnewline
{\footnotesize{}$\delta=1/32$} &  & {\footnotesize{}$\tau^{e}=2.9$ hours} &  & {\footnotesize{}$4.1\times\tau^{e}$} &  & {\footnotesize{}$7.0\times\tau^{e}$} &  & {\footnotesize{}$13.6\times\tau^{e}$} &  & {\footnotesize{}$35.4\times\tau^{e}$}\tabularnewline
\hline 
\end{tabular}
\par\end{centering}
\end{singlespace}
\smallskip

{\footnotesize{}This table reports the computational runtime w.r.t.
control dimension of approximating conditional expectations for one
time step under different sizes of mesh and different regression bases.
A CRRA investment style with $\gamma=10$ over a 6-month horizon ($N=6$
months) is considered. For ease of comparison, the results are reported
as multiple of the fastest method (local regression). }{\footnotesize\par}
\end{table}

\subsection{Mesh size \label{subsec:Mesh-size}}

We then test the sensitivity of the accuracy (Table \ref{tab:mesh})
and efficiency (Table \ref{tab:runtime}) with respect to the mesh
size of the control grid. Table \ref{tab:mesh} shows that a coarse
grid ($\delta=1/8$) combined with local control regression is able
to produce accurate results: the errors in CER and initial portfolio
allocation are negligible compared to a fine grid ($\delta=1/32$). 

Table \ref{tab:runtime} reports the runtime of one time step iteration
of the backward dynamic programming for different control meshes.
The main message coming from these results is that although we still
do not completely get rid of the curse of dimensionality, the presented
technique allows us to solve portfolio optimization problems of much
greater size (more than ten risky assets) than what is usually considered
in the literature (less than three risky assets).

It is important to note that, in order to properly compare the runtime,
the code is written based on naive ``for loop'' for each Monte Carlo
path and the computational speed can be greatly improved by vectorization
or by using lower level programming languages. In particular, the
code is implemented in Python 3.4.3 on a single processor Intel Core
i7 2.2 GHz.
\begin{table}[H]
\begin{singlespace}
\begin{centering}
\caption{CER and initial optimal allocation with different mesh sizes\label{tab:mesh}}
\smallskip{\footnotesize{}}%
\begin{tabular}{>{\raggedright}m{12mm}l>{\raggedright}m{12mm}llll>{\raggedright}p{0.5mm}lll}
 &  &  &  & {\footnotesize{}CER} &  &  &  & {\footnotesize{}Initial weights $\boldsymbol{\alpha}$} &  & \tabularnewline
\hline 
 &  &  &  & {\footnotesize{}$\gamma\!=\!5$} & {\footnotesize{}$\gamma\!=\!10$} & {\footnotesize{}$\gamma\!=\!15$} &  & {\footnotesize{}$\gamma\!=\!5$} & {\footnotesize{}$\gamma\!=\!10$} & {\footnotesize{}$\gamma\!=\!15$}\tabularnewline
\hline 
\multirow{5}{12mm}{{\footnotesize{}$N=3$}} &  & {\footnotesize{}$\delta=1/2$} &  & {\footnotesize{}$66.6$} & {\footnotesize{}$57.5$} & {\footnotesize{}$47.3$} &  & {\footnotesize{}$0.51$, $0.37$, $0.12$} & {\footnotesize{}$0.78$, $0.22$, $0.00$} & {\footnotesize{}$1.00$, $0.00$, $0.00$}\tabularnewline
 &  & {\footnotesize{}$\delta=1/4$} &  & {\footnotesize{}$66.6$} & {\footnotesize{}$57.6$} & {\footnotesize{}$47.8$} &  & {\footnotesize{}$0.41$, $0.31$, $0.18$} & {\footnotesize{}$0.65$, $0.22$, $0.13$} & {\footnotesize{}$0.95$, $0.05$, $0.00$}\tabularnewline
 &  & {\footnotesize{}$\delta=1/8$} &  & {\footnotesize{}$66.4$} & {\footnotesize{}$57.8$} & {\footnotesize{}$48.0$} &  & {\footnotesize{}$0.42$, $0.32$, $0.16$} & {\footnotesize{}$0.64$, $0.23$, $0.13$} & {\footnotesize{}$0.92$, $0.08$, $0.00$}\tabularnewline
 &  & {\footnotesize{}$\delta=1/16$} &  & {\footnotesize{}$66.6$} & {\footnotesize{}$57.8$} & {\footnotesize{}$47.8$} &  & {\footnotesize{}$0.42$, $0.33$, $0.15$} & {\footnotesize{}$0.65$, $0.24$, $0.11$} & {\footnotesize{}$0.92$, $0.08$, $0.00$}\tabularnewline
 &  & {\footnotesize{}$\delta=1/32$} &  & {\footnotesize{}$66.7$} & {\footnotesize{}$57.8$} & {\footnotesize{}$48.0$} &  & {\footnotesize{}$0.41$, $0.32$, $0.17$} & {\footnotesize{}$0.64$, $0.26$, $0.10$} & {\footnotesize{}$0.92$, $0.08$, $0.00$}\tabularnewline
 &  &  &  &  &  &  &  &  &  & \tabularnewline
\multirow{5}{12mm}{{\footnotesize{}$N=6$}} &  & {\footnotesize{}$\delta=1/2$} &  & {\footnotesize{}$80.8$} & {\footnotesize{}$64.9$} & {\footnotesize{}$49.7$} &  & {\footnotesize{}$0.65$, $0.27$, $0.08$} & {\footnotesize{}$0.85$, $0.15$, $0.00$} & {\footnotesize{}$1.00$, $0.00$, $0.00$}\tabularnewline
 &  & {\footnotesize{}$\delta=1/4$} &  & {\footnotesize{}$80.8$} & {\footnotesize{}$65.0$} & {\footnotesize{}$50.2$} &  & {\footnotesize{}$0.58$, $0.31$, $0.11$} & {\footnotesize{}$0.77$, $0.18$, $0.05$} & {\footnotesize{}$1.00$, $0.00$, $0.00$}\tabularnewline
 &  & {\footnotesize{}$\delta=1/8$} &  & {\footnotesize{}$80.6$} & {\footnotesize{}$65.1$} & {\footnotesize{}$50.5$} &  & {\footnotesize{}$0.57$, $0.31$, $0.12$} & {\footnotesize{}$0.75$, $0.20$, $0.05$} & {\footnotesize{}$0.91$, $0.09$, $0.00$}\tabularnewline
 &  & {\footnotesize{}$\delta=1/16$} &  & {\footnotesize{}$80.7$} & {\footnotesize{}$65.2$} & {\footnotesize{}$50.5$} &  & {\footnotesize{}$0.55$, $0.32$, $0.13$} & {\footnotesize{}$0.74$, $0.20$, $0.06$} & {\footnotesize{}$0.92$, $0.08$, $0.00$}\tabularnewline
 &  & {\footnotesize{}$\delta=1/32$} &  & {\footnotesize{}$80.8$} & {\footnotesize{}$65.2$} & {\footnotesize{}$50.5$} &  & {\footnotesize{}$0.55$, $0.32$, $0.13$} & {\footnotesize{}$0.71$, $0.22$, $0.07$} & {\footnotesize{}$0.92$, $0.08$, $0.00$}\tabularnewline
 &  &  &  &  &  &  &  &  &  & \tabularnewline
\multirow{5}{12mm}{{\footnotesize{}$N=12$}} &  & {\footnotesize{}$\delta=1/2$} &  & {\footnotesize{}$86.5$} & {\footnotesize{}$66.7$} & {\footnotesize{}$51.0$} &  & {\footnotesize{}$0.87$, $0.13$, $0.00$} & {\footnotesize{}$1.00$, $0.00$, $0.00$} & {\footnotesize{}$1.00$, $0.00$, $0.00$}\tabularnewline
 &  & {\footnotesize{}$\delta=1/4$} &  & {\footnotesize{}$86.7$} & {\footnotesize{}$67.1$} & {\footnotesize{}$51.0$} &  & {\footnotesize{}$0.81$, $0.19$, $0.00$} & {\footnotesize{}$0.90$, $0.10$, $0.00$} & {\footnotesize{}$0.86$, $0.14$, $0.00$}\tabularnewline
 &  & {\footnotesize{}$\delta=1/8$} &  & {\footnotesize{}$86.5$} & {\footnotesize{}$67.2$} & {\footnotesize{}$51.2$} &  & {\footnotesize{}$0.82$, $0.18$, $0.00$} & {\footnotesize{}$0.89$, $0.11$, $0.00$} & {\footnotesize{}$0.88$, $0.12$, $0.00$}\tabularnewline
 &  & {\footnotesize{}$\delta=1/16$} &  & {\footnotesize{}$86.6$} & {\footnotesize{}$67.4$} & {\footnotesize{}$51.1$} &  & {\footnotesize{}$0.81$, $0.19$, $0.00$} & {\footnotesize{}$0.87$, $0.13$, $0.00$} & {\footnotesize{}$0.88$, $0.12$, $0.00$}\tabularnewline
 &  & {\footnotesize{}$\delta=1/32$} &  & {\footnotesize{}$86.8$} & {\footnotesize{}$67.4$} & {\footnotesize{}$51.1$} &  & {\footnotesize{}$0.81$, $0.19$, $0.00$} & {\footnotesize{}$0.87$, $0.13$, $0.00$} & {\footnotesize{}$0.88$, $0.12$, $0.00$}\tabularnewline
\hline 
\end{tabular}{\footnotesize\par}
\par\end{centering}
\end{singlespace}
\smallskip

{\footnotesize{}This table compares the monthly adjusted CER (in basis
points) and the initial optimal portfolio allocation using different
sizes of control mesh ($\delta=1/2,1/4,1/8,1/16,1/32$) under different
investment horizons ($N=3,6,12$ months) and different risk-aversion
parameters of the CRRA utility ($\gamma=5,10,15$).}{\footnotesize\par}
\end{table}

\begin{table}[H]
\caption{Computational runtime with different control dimensions\label{tab:runtime}}
\smallskip
\begin{singlespace}
\begin{centering}
\begin{tabular}{ccccccccccc}
 &  & {\footnotesize{}$\delta=1/2$} &  & {\footnotesize{}$\delta=1/4$} &  & {\footnotesize{}$\delta=1/8$} &  & {\footnotesize{}$\delta=1/16$} &  & {\footnotesize{}$\delta=1/32$}\tabularnewline
\hline 
{\footnotesize{}$d=2$} &  & {\footnotesize{}$19\text{ secs}$} &  & {\footnotesize{}$26\text{ secs}$} &  & {\footnotesize{}$48\text{ secs}$} &  & {\footnotesize{}$1.6\text{ mins}$} &  & {\footnotesize{}$3.2\text{ mins}$}\tabularnewline
{\footnotesize{}$d=3$} &  & {\footnotesize{}$43\text{ secs}$} &  & {\footnotesize{}$1.5\text{ mins}$} &  & {\footnotesize{}$5.2\text{ mins}$} &  & {\footnotesize{}$24.1\text{ mins}$} &  & {\footnotesize{}$2.9\text{ hrs}$}\tabularnewline
{\footnotesize{}$d=4$} &  & {\footnotesize{}$1.4\text{ mins}$} &  & {\footnotesize{}$4.5\text{ mins}$} &  & {\footnotesize{}$29.4\text{ mins}$} &  & {\footnotesize{}$8.6\text{ hrs}$} &  & {\footnotesize{}N\textbackslash A}\tabularnewline
{\footnotesize{}$d=5$} &  & {\footnotesize{}$2.4\text{ mins}$} &  & {\footnotesize{}$11.8\text{ mins}$} &  & {\footnotesize{}$2.9\text{ hrs}$} &  & {\footnotesize{}N\textbackslash A} &  & {\footnotesize{}N\textbackslash A}\tabularnewline
{\footnotesize{}$d=6$} &  & {\footnotesize{}$3.9\text{ mins}$} &  & {\footnotesize{}$37.2\text{ mins}$} &  & {\footnotesize{}$13.8\text{ hrs}$} &  & {\footnotesize{}N\textbackslash A} &  & {\footnotesize{}N\textbackslash A}\tabularnewline
{\footnotesize{}$d=7$} &  & {\footnotesize{}$6.1\text{ mins}$} &  & {\footnotesize{}$1.3\text{ hrs}$} &  & {\footnotesize{}N\textbackslash A} &  & {\footnotesize{}N\textbackslash A} &  & {\footnotesize{}N\textbackslash A}\tabularnewline
{\footnotesize{}$d=8$} &  & {\footnotesize{}$9.1\text{ mins}$} &  & {\footnotesize{}$2.2\text{ hrs}$} &  & {\footnotesize{}N\textbackslash A} &  & {\footnotesize{}N\textbackslash A} &  & {\footnotesize{}N\textbackslash A}\tabularnewline
{\footnotesize{}$d=9$} &  & {\footnotesize{}$14.0\text{ mins}$} &  & {\footnotesize{}$3.5\text{ hrs}$} &  & {\footnotesize{}N\textbackslash A} &  & {\footnotesize{}N\textbackslash A} &  & {\footnotesize{}N\textbackslash A}\tabularnewline
{\footnotesize{}$d=10$} &  & {\footnotesize{}$19.6\text{ mins}$} &  & {\footnotesize{}$6.4\text{ hrs}$} &  & {\footnotesize{}N\textbackslash A} &  & {\footnotesize{}N\textbackslash A} &  & {\footnotesize{}N\textbackslash A}\tabularnewline
\hline 
\end{tabular}
\par\end{centering}
\end{singlespace}
\smallskip{\footnotesize{}This table reports the computational runtime
w.r.t. control dimension of performing one time step conditional expectation
approximation under different sizes of mesh ($\delta=1/2,1/4,1/8,1/16,1/32$).
``N\textbackslash A'' indicates the situation where the program
takes more than 24 hours to run. }{\footnotesize\par}
\end{table}

\section{Conclusion\label{sec:Conclusion}}

This paper designs an improved LSMC method for solving stochastic
control problems. This LSMC modification can be summarized as follows:
we first estimate the value functions on a discrete grid of controls,
then generalize the discrete estimates by local control regression,
and finally determine the optimal (continuous) control by an adaptive
refinement algorithm. We apply the method to a dynamic multivariate
portfolio optimization problem in the presence of transaction costs,
liquidity costs and market impact. We numerically show that the local
regression of continuation estimates with respect to portfolio weights
is more accurate and more efficient than the classical global control
regression. We show that the combination of local control regression
and adaptive grids significantly improves the computational efficiency
without sacrificing accuracy, and that accurate results can be obtained
from very coarse grids. 

\section*{Acknowledgment}

Part of this research is funded by RiskLab Australia.

\bibliographystyle{chicago}
\bibliography{bib_portfolio,bib_grid}

\begin{thebibliography}{}

\bibitem[\protect\citeauthoryear{Brandt, Goyal, Santa-Clara, and Stroud}{Brandt
  et~al.}{2005}]{Brandt05}
Brandt, M., A.~Goyal, P.~Santa-Clara, and J.~Stroud (2005).
\newblock A simulation approach to dynamic portfolio choice with an application
  to learning about return predictability.
\newblock {\em Review of Financial Studies\/}~{\em 18\/}(3).

\bibitem[\protect\citeauthoryear{Broadie and Shen}{Broadie and
  Shen}{2016}]{Broadie2016}
Broadie, M. and W.~Shen (2016).
\newblock High-dimensional portfolio optimization with transaction costs.
\newblock {\em International Journal of Theoretical and Applied Finance\/}~{\em
  19\/}(4), 1650025.

\bibitem[\protect\citeauthoryear{Carriere}{Carriere}{1996}]{Carriere1996}
Carriere, J. (1996).
\newblock Valuation of the early-exercise price for options using simulations
  and nonparametric regression.
\newblock {\em Insurance: Mathematics and Economics\/}~{\em 19\/}(1), 19--30.

\bibitem[\protect\citeauthoryear{Cong and Oosterlee}{Cong and
  Oosterlee}{2016a}]{Cong2016}
Cong, F. and C.~W. Oosterlee (2016a).
\newblock Multi-period mean-variance portfolio optimization based on {M}onte
  {Carlo} simulation.
\newblock {\em Journal of Economic Dynamics and Control\/}~{\em 64\/}(1),
  23--38.

\bibitem[\protect\citeauthoryear{Cong and Oosterlee}{Cong and
  Oosterlee}{2016b}]{Cong2016b}
Cong, F. and C.~W. Oosterlee (2016b).
\newblock On pre-commitment aspects of a time-consistent strategy for a
  mean-variance investor.
\newblock {\em Journal of Economic Dynamics and Control\/}~{\em 70\/}(1),
  178--193.

\bibitem[\protect\citeauthoryear{Cong and Oosterlee}{Cong and
  Oosterlee}{2017}]{Cong2017}
Cong, F. and C.~W. Oosterlee (2017).
\newblock Accurate and robust numerical methods for the dynamic portfolio
  management problem.
\newblock {\em Computational Economics\/}~{\em 49\/}(3), 433--458.

\bibitem[\protect\citeauthoryear{Denault, Delage, and Simonato}{Denault
  et~al.}{2017}]{Denault17b}
Denault, M., E.~Delage, and J.-G. Simonato (2017).
\newblock Dynamic portfolio choice: a simulation-and-regression approach.
\newblock {\em Optimization and Engineering\/}~{\em 18\/}(2), 369--406.

\bibitem[\protect\citeauthoryear{Denault and Simonato}{Denault and
  Simonato}{2017}]{Denault2017}
Denault, M. and J.-G. Simonato (2017).
\newblock Dynamic portfolio choices by simulation-and-regression: Revisiting
  the issue of value function vs portfolio weight recursions.
\newblock {\em Computers and Operations Reseach\/}~{\em 79}, 174--189.

\bibitem[\protect\citeauthoryear{Diris, Palm, and Schotman}{Diris
  et~al.}{2015}]{Diris15}
Diris, B., F.~Palm, and P.~Schotman (2015).
\newblock Long-term strategic asset allocation: an out-of-sample evaluation.
\newblock {\em Management Science\/}~{\em 61\/}(9), 2185--2202.

\bibitem[\protect\citeauthoryear{Garlappi and Skoulakis}{Garlappi and
  Skoulakis}{2010}]{Garlappi2010}
Garlappi, L. and G.~Skoulakis (2010).
\newblock Solving consumption and portfolio choice problems: The state variable
  decomposition method.
\newblock {\em Review of Financial Studies\/}~{\em 23}, 3346--3400.

\bibitem[\protect\citeauthoryear{Glasserman and Yu}{Glasserman and
  Yu}{2002}]{Glasserman2002}
Glasserman, P. and B.~Yu (2002).
\newblock Simulation for {A}merican options: Regression now or regression
  later?
\newblock {\em Monte Carlo and Quasi-Monte Carlo Methods 2002\/}, 213--226.

\bibitem[\protect\citeauthoryear{Kharroubi, Langren\'e, and Pham}{Kharroubi
  et~al.}{2014}]{Kharroubi2014}
Kharroubi, I., N.~Langren\'e, and H.~Pham (2014).
\newblock A numerical algorithm for fully nonlinear {HJB} equations: an
  approach by control randomization.
\newblock {\em Monte Carlo Methods and Applications\/}~{\em 20\/}(2), 145--165.

\bibitem[\protect\citeauthoryear{Koijen, Nijman, and Werker}{Koijen
  et~al.}{2010}]{Koijen10}
Koijen, R., T.~Nijman, and B.~Werker (2010).
\newblock When can life cycle investors benefit from time-varying bond risk
  premia.
\newblock {\em The Review of Financial Studies\/}~{\em 23\/}(2), 741--780.

\bibitem[\protect\citeauthoryear{Koijen, Rodr\'iguez, and Sbuelz}{Koijen
  et~al.}{2009}]{Koijen09}
Koijen, R., J.~C. Rodr\'iguez, and A.~Sbuelz (2009).
\newblock Momentum and mean reversion in strategic asset allocation.
\newblock {\em Management Science\/}~{\em 55\/}(7), 1199--1213.

\bibitem[\protect\citeauthoryear{Longstaff and Schwartz}{Longstaff and
  Schwartz}{2001}]{Longstaff2001}
Longstaff, F. and E.~Schwartz (2001).
\newblock Valuing {A}merican options by simulation: A simple least-squares
  approach.
\newblock {\em Review of Financial Studies\/}~{\em 14\/}(1), 681--692.

\bibitem[\protect\citeauthoryear{Shen, Pelsser, and Schotman}{Shen
  et~al.}{2014}]{Shen14}
Shen, S., A.~Pelsser, and P.~Schotman (2014).
\newblock Robust long-term interest rate risk hedging in incomplete bond
  markets.
\newblock Technical report, NETSPAR.

\bibitem[\protect\citeauthoryear{Tian, Rood, and Oosterlee}{Tian
  et~al.}{2013}]{Tian2013}
Tian, Y., R.~Rood, and C.~Oosterlee (2013).
\newblock Efficient portfolio valuation incorporating liquidity risk.
\newblock {\em Quantitative Finance\/}~{\em 13\/}(10), 1575--1586.

\bibitem[\protect\citeauthoryear{Tsitsiklis and Van~Roy}{Tsitsiklis and
  Van~Roy}{2001}]{Tsitsiklis01}
Tsitsiklis, J. and B.~Van~Roy (2001).
\newblock Regression methods for pricing complex {A}merican-style options.
\newblock {\em IEEE Transactions on Neural Networks\/}~{\em 12\/}(4), 694--703.

\bibitem[\protect\citeauthoryear{Van~Binsbergen and Brandt}{Van~Binsbergen and
  Brandt}{2007}]{vanBinsbergen2007}
Van~Binsbergen, J.~H. and M.~Brandt (2007).
\newblock Solving dynamic portfolio choice problems by recursing on optimized
  portfolio weights or on the value function?
\newblock {\em Computational Economics\/}~{\em 29}, 355--367.

\bibitem[\protect\citeauthoryear{Zhang, Langren\'e, Tian, Zhu, Klebaner, and
  Hamza}{Zhang et~al.}{2018}]{Zhang2018}
Zhang, R., N.~Langren\'e, Y.~Tian, Z.~Zhu, F.~Klebaner, and K.~Hamza (2018).
\newblock Dynamic portfolio optimization with liquidity cost and market impact:
  a simulation-and-regression approach.
\newblock To appear in Quantitative Finance.

\end{thebibliography}

\end{document}